# Decomposed Interval Type-2 Fuzzy Systems with Application to Inverted Pendulum

Sherif M. Abuelenin

Department of Electrical Engineering, Faculty of Engineering
Port-Said University
Port-Fouad, Port-Said, Egypt
sherif217@ieee.org

*Abstract*—This article introduces the idea of decomposition of interval Type-2 fuzzy logic system into two parallel type-1 fuzzy systems. This decomposition avoids the problems associated with type-reduction techniques normally needed in type-2 fuzzy systems. Next, we compare the performance of a decomposed type-2 controller to the performance of a type-1 controller in stabilizing an inverted pendulum.

## I. Introduction

Fuzzy set theory was introduced by Zadeh [1] in 1965. Since then, fuzzy logic systems have been applied in many areas, including modeling and control. Fuzzy controllers have shown superiority over other controllers [2], especially in situations where the controlled systems cannot be analytically modeled. Traditional fuzzy logic systems are referred to as Type-1 fuzzy logic systems (T1 FLSs). They were criticized, regardless their name, of not being able to model and account for uncertainties. Such uncertainties may come from [3]:

a) Different meanings of words -used in fuzzy rule base-to different people

b) Experts designing the fuzzy system having different opinions

c) Noisy training data

d) Existence of noise in measurements of FLS inputs.

To deal with this criticism, in 1975 Zadeh [4] introduced type-2 fuzzy sets [5]. A type-2 fuzzy set (T2 FS) is a three dimensional fuzzy set. In which the *primary* fuzzy set is characterized by membership grades that are not crisp numbers, but rather fuzzy sets themselves; those are called secondary membership functions [6], [7]. A type-2 fuzzy set has a footprint of uncertainty that represents uncertainties in the definition of the fuzzy set. When the secondary MF is a unit interval for all the points in the primary membership this set is called an interval type-2 fuzzy set (IT2 FS) [8].

A fuzzy system that has type-2 fuzzy sets in its antecedent or consequent part is referred to as a type-2 fuzzy logic system (T2 FLS). It has been demonstrated that type-2 FLSs are capable of dealing with all such uncertainties [3]. And it was shown that T2 FLSs can outperform T1 FLSs in a many fields of application [9], including control [10]. Fuzzy systems that utilize IT2 FSs are called interval type-2 fuzzy logic systems (IT2 FLSs) [11].

The output of an IT2 FLS is an IT2 FS. It requires the conversion to T1 FS before the final crisp output can be calculated. This process is called type-reduction [6]. Several methods have been introduced for type reduction; each has its own deficiencies. In this paper we introduce the method of IT2 FLS decomposition an alternative to type-reduction.

The rest of this paper is organized as follows; we introduce IT2 fuzzy sets and systems in section II, and we discuss type reduction methods and their limitations in section III. In section IV we introduce the method of decomposition of ITS FLSs as an alternative to type-reduction. In section V we show an example of a decomposed IT2 fuzzy logic controller (FLC) and compare its performance to a T1 FLC. Conclusions are provided in section VI.

## II. Interval Type-2 Fuzzy Systems

### A. Interval type-2 fuzzy sets

An example of an IT2 FS is shown in Fig. 1. The fuzzy set shown has certain mean and uncertainty in the end points. Other T2 sets may include uncertainties in mean, standard deviation, etc. Note that the third dimension of the IT2 can be ignored in analysis because it conveys no new information, and uncertainty in the primary memberships of an IT2 FS consists of a region bounded between the upper and lower membership functions. This region is called the footprint of uncertainty (FOU). [12]

### B. Interval type-2 fuzzy systems

As Fig. 2 shows; the major difference between a T1 and IT2 system is that the latter uses at least one IT2 fuzzy set. Hence the output of the inference mechanism contains IT2 fuzzy sets. This requires an additional step of type reduction (mapping a T2 FS into a T1 FS) before the final (crisp) output can be calculated. In the following section we discuss common methods of type-reduction, and their limitations.

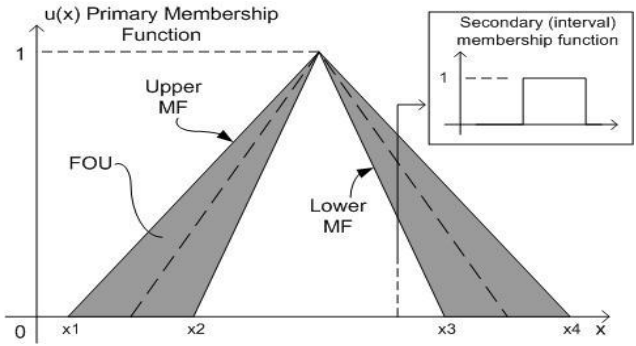

Figure 1. Example of an IT2 Fuzzy Set, with the secondary membership function shown in the inset.

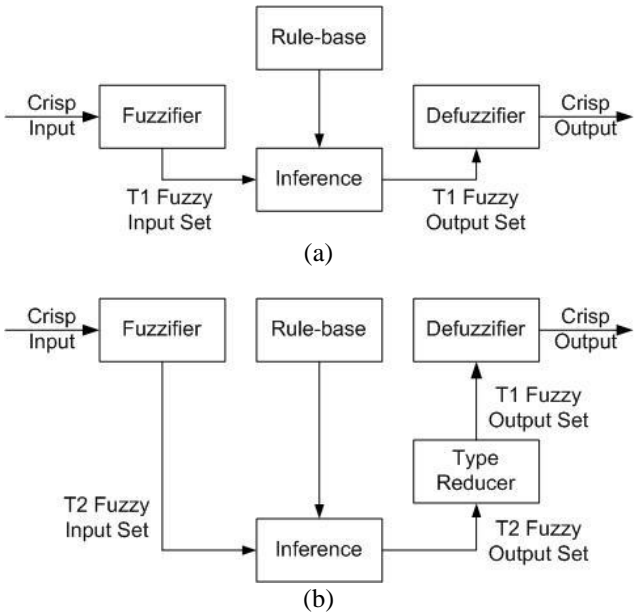

Figure 2. (a) Type-1 fuzzy logic system, (b) Type-2 fuzzy logic system

## III. TYPE REDUCTION

As shown in Fig. 2, defuzzification of a type-2 fuzzy may be considered a two-stage process. The first stage is type-reduction that produces a type-reduced set; a type-1 fuzzy set that is a fuzzy representation of the centroid of the type-2 fuzzy set. This set can then be defuzzified to give a crisp number [13]. Type-reduction is an extension of T1 defuzzification, by applying the extension principle [14] to a selected method of defuzzification. It transforms a T2 fuzzy set into a T1 fuzzy set [3]. In IT2 systems the type-reduced set is an interval set that is determined by two end points.

A popular type reduction method is to compute the centroid of an IT2 FS. Centroid type-reduction for IT2 FLSs was developed by Karnik and Mendel [14]. Because of the iterative natures of the KM algorithms, they may produce a computational bottleneck in real-time applications. Wu and Mendel [15]-[17] proposed enhanced KM (EKM) algorithms to reduce the computational cost of the standard KM algorithms [3], but, they were still iterative.

Other methods were introduced for T2 centroid computation. The collapsing method [18] is an approximation to the KM algorithm. The method of uncertainty bound approximates type reduction of IT2 FLSs [3] by computing four uncertainty bounds using the centroids of M consequent IT2 FSs, and then an approximate type-reduced set is found before an approximate defuzzified output is calculated [6]. This method is rather complex [19] and provides an approximate output. More recently Centroid-Flow algorithm was introduced to reduce the computation time in comparison to both KM and EKM algorithms [20]. A comparison of common type-reduction algorithms is introduced in [21]. In real-time applications, especially control systems, the iterative type reduction (TR) methods were not favored because they have no closed-form solution. Also uncertainty bounds were difficult to use in analyses of control systems. Much research occurred and is still occurring on ways to bypass TR so that IT2 FL controllers can be used in real time [19].

Coupland and John [13], [22]-[24] proposed an alternative geometric defuzzification method to type-reduction for the defuzzification of IT2 FSs using computational geometry. In this method, the FOU of the three dimensional IT2 FS was approximated by using regular geometric shapes, mostly triangles. Operations such as union and intersection are carried out using methods from computational geometry. Type-reduction is by-passed by directly using one coordinate of the geometric centroid for the defuzzified value of the T2 FS [19]. The final consequent IT2 set is defuzzified by calculating the geometric center of its FOU. To find the geometric center, the FOU is converted to a set of closed nonintersecting polygons, and then the weighted average of the polygons is calculated.

The geometric centroid (GC) introduced good results that were close to the type-reduced centroid. It by-passed the iterative type-reduction methods and the complexities of using uncertainty bounds, especially in real time control applications. However, the GC has a limitation on the form of fuzzy sets being used [25]. Some fuzzy membership functions are complex to represent in geometrical forms. And the union and intersection operations needed to be carried out using computational geometry.

Following on this method, we describe in the following section the concept of decomposition of an IT2 FLS into two parallel T1 FLSs. The final output of the system can be found by computing the geometric centroid of the aggregated output of both controllers. The output in this case is equivalent to the output computed by CG method, but utilizing only T1 operations and T1 membership functions, which simplifies the analysis, and permits the use of any shape of T1 membership functions.

## IV. DECOMPOSITION OF IT2 SYSTEMS

The proposed approach of decomposition of an IT2 FLS is illustrated in Fig. 3. The idea is to perform the fuzzy analysis in two separate paths. The crisp input is fuzzified using the upper membership functions in one path and the lower membership functions on the other path. The T1 fuzzy outputs of both controllers are computed separately using the inference engines and the rule base. The last step is to combine both sets to find the FOU for the output membership function.

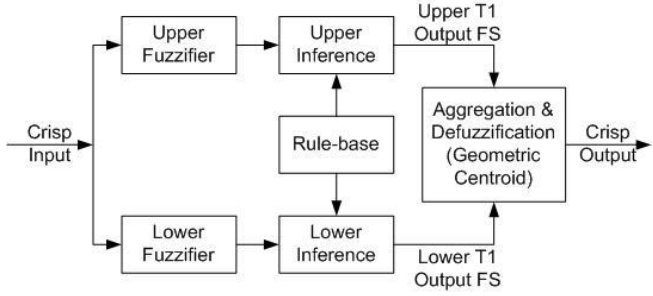

Figure 3. Structure of decomposed IT2 fuzzy logic system.

The geometric center of the FOU of an IT2 fuzzy set is found. Unlike the GC method which requires the conversion of the FOU to a set of closed nonintersecting polygons, we find the crisp output by finding the center-of-area (centroid) of the FOU using geometrical means. This is simply done by computing the centroid of and the area under both the upper and lower membership functions, and then computing the centroid using (1). Note that we consider the area under the lower output MF to be a negative area.

$$y = \frac{c_U A_U - c_L A_L}{A_U - A_L} \quad (1)$$

Where $A_U$ is the area under the upper output membership function, $A_L$ is the area under the lower output membership function, $c_U$ is the centroid of the upper output membership function, and $c_L$ is the centroid of the lower output membership function. Both centroids can be computed using T1 centroid computation methods.

If both output sets are identical, then there is no uncertainty and the system reduces to an equivalent type-1 system. The crisp output then is simple equivalent to $c_U$ (which will also be equal to $c_L$ in this case).

## V. APPLICATION TO THE INVERTED PENDULUM PROBLEM

In order to examine the performance of the proposed method, we designed a decomposed IT2 FLC for an inverted pendulum. Designing a controller for balancing an inverted pendulum mounted on a cart is a well known control problem. In this section we show the response of the introduced decomposed IT2 controller and compare its performance with that of a corresponding T1 FLC. All simulations are performed using MATLAB® and SIMULINK®.

### A. The inverted Pendulum

An inverted pendulum is a pendulum that is mounted on a cart in the way shown in Fig. 4. It is balanced by applying dynamic force $f$ to the cart. Feedback control is used as shown in Fig. 5. We use the model described by (2) and (3) for the inverted pendulum [26].

$$\ddot{y} = \frac{g\sin(y) + \cos(y)\left[\dfrac{-\bar{f} - 0.25\dot{y}^2 \sin(y)}{1.5}\right]}{\left[\dfrac{2}{3} - \dfrac{1}{6}\cos^2(y)\right]} \quad (2)$$

$$\dot{\bar{f}} = -100\bar{f} + 100f \quad (3)$$

Where, $f$ is the applied force in Newton, $y$ is the angular position in radians, and g is the gravity of earth.

In order to balance the inverted pendulum we use closed-loop feedback fuzzy control, as shown in Fig. 5. It is common to use the error signal and its derivative as inputs to the controller. We use both T1 and Decomposed T2 FLCs to balance the system and compare their performances. For both controllers, we use the same number of membership functions (three) to fuzzify each input. We also use the same rule-base for inference in both cases. Minimum was used for AND, and implication, and maximum was used for aggregation of all the fuzzy systems utilized.

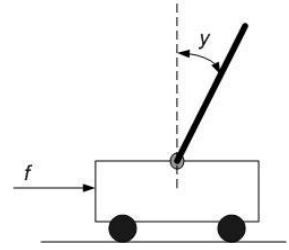

Figure 4. Inverted pendulum system.

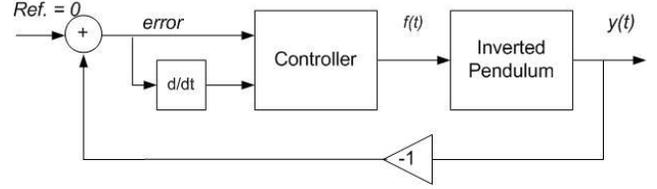

Figure 5. Feedback controller for inverted pendulum.

### B. Type-1 Fuzzy Control

The rule-base for controlling the inverted pendulum is shown in Fig. 6. For simplicity we use only three triangular shaped membership functions for each of the input and output variables. The T1 fuzzy input and output membership functions are shown in Fig. 7. a, b, and c. For each variable the three linguistic variables are named *N*, *Z*, and *P* (negative, zero, and positive, respectively). The inputs to the fuzzy controllers are saturated to a maximum and minimum of pi/4 and –pi/4 respectively.

### C. Decomposed IT2 Fuzzy Control

The control is now performed using the proposed decomposed controller. Input membership functions are changed to IT2 membership functions. This is done by blurring out the T1 fuzzy sets []. We assume no change in the mean values, and uncertainty in the width of the membership functions. We used uncertainty of pi/16 in each direction for all membership functions of the input. The output membership functions were kept T1 functions. Fig. 8. shows the used input IT2 membership functions.

the input of the controller; the result is shown in Fig. 10. It is clear from both figures that the proposed controller outperforms the T1 controller. The response of the system with the proposed controllers is faster than that of the T1 controller. The effect of noise is also less significant with the decomposed IT2 controller.

| Force 'f' | | Error derivative | | |
|---|---|---|---|---|
| | | N | Z | P |
| Error | N | P | P | Z |
| | Z | P | Z | N |
| | P | Z | N | N |

Figure 6. Rule-base for fuzzy control.

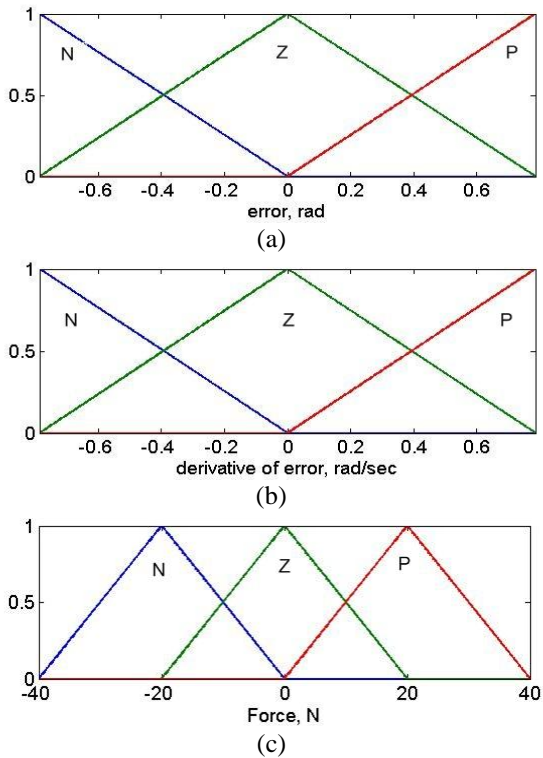

Figure 7. T1 fuzzy membership functions: (a) error; (b) error derivative; (c) force.

*D. Simulation Results*

We performed simulation of the inverted pendulum system with both using SIMULINK under similar simulation parameters. Discrete T1 and IT2 controllers were programmed as M-files. The system was simulated with each of the two controllers under the same initial conditions; initial angular position of *0.1 rad*, and initial angular velocity equals zero.

Fig. 9. shows the response of both systems with no added noise. We repeated the simulation, adding Gaussian noise to

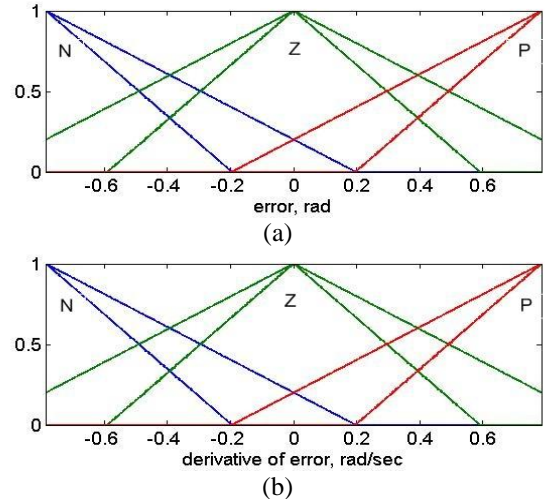

Figure 8. T2 fuzzy membership functions: (a) error; (b) error derivative.

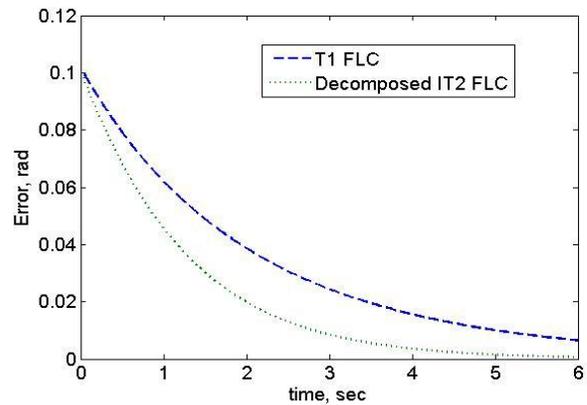

Figure 9. Inverted pendulum response with T1 and decomposed IT2 controllers.

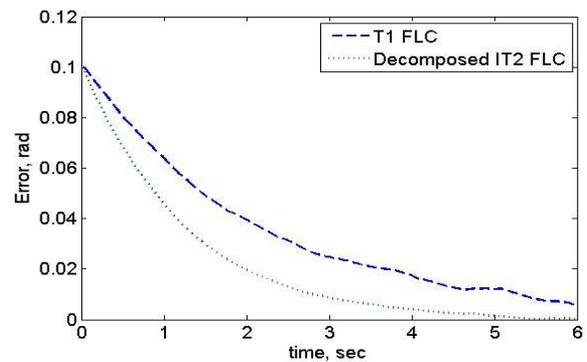

Figure 10. Inverted pendulum response with T1 and decomposed IT2 controllers in presence of noise.

## VI. Conclusions

We discussed the concept of decomposition of interval type-2 fuzzy logic system as an alternative to type-reduction methods. It avoids the problems associated with type-reduction methods such as computational complexities, and the iterative nature of some common type-reduction algorithms, enabling easier implementation in real-time applications. Because the method is based completely on type-1 operations, It allows the use of any form of input and output fuzzy membership functions.

The decomposition method extends Coupland's geometric defuzzification method to use only type-1 analysis in evaluating the IT2 FLS. By decomposing an IT2 system into two T1 systems, we can use type-1 fuzzy operations and membership functions for the fuzzification and inference steps. The two resulting output T1 fuzzy membership functions are combined producing the FOU of an equivalent IT2 set, which is defuzzified using geometric centroid method to compute the output of the system.

The introduced method was used to control an inverted pendulum. It showed better performance in comparison to T1 control, in both noise-free and noisy conditions. All simulations were performed using MATLAB® and SIMULINK®.